\documentstyle[12pt]{article}

\def\ga{\gamma}
\def\pd{\partial}
\newcommand\bb{\begin{equation}}
\newcommand\bba{\begin{eqnarray}}
\newcommand\ee{\end{equation}}
\newcommand\eea{\end{eqnarray}}

\begin{document}
\begin{titlepage}
%
\centerline{\Huge On the Bound States in a Non-linear}
\vskip 3.5mm\centerline{\Huge Quantum Field Theory of a Spinor
Field}
\vskip 2mm\centerline{\Huge with Higher
Derivatives\Large \footnote{Supported in
part by BNSF under contract $\Phi$-401}}
\vskip 15mm\centerline{\Large\it{ A.D. Mitov,\footnote{
e-mail:amitov@inrne.acad.bg}   M.N. Stoilov \footnote{
e-mail:mstoilov@inrne.acad.bg} and  D.Ts. Stoyanov \footnote
{e-mail:dstoanov@inrne.acad.bg}}}
\vskip 10mm\centerline{\it Institute for Nuclear Research and 
Nuclear Energy}\centerline{\it Boul. Tsarigradsko Chausse\'e 72, 
Sofia 1784, Bulgaria}
\vskip 10mm\centerline{11th July 1997}
\vskip 25mm\centerline{\bf Abstract}\vskip 2mm
We consider a model of quantum field theory with higher
derivatives for a spinor field with $(\bar\varphi\varphi)^2$
selfinteraction. With the help of the Bethe--Salpeter
equation we study the problem of the two particle bound states 
in the "chain" approximation.
The existence of a scalar bound state is established.

\textheight 196mm
\end{titlepage}

\voffset 0mm
\vfill\eject\newpage
\textheight 196mm
\section{Introduction}\vskip 5mm

One of the basic problems in Quantum Field Theory (QFT)
is related to the presence of divergences in the $S$-matrix
elements.
The reason for their appearance is well known: the
propagators do not tend sufficiently quick to zero at large
momenta. In standard QFT where one deals with equations of motion
for Fermi (Boze) fields which are of first (second) order in the
derivatives, the only solution of this problem is given by the
renormalization method.
The core  of this method consists in the following -- due to
the arbitrariness in the $S$-matrix \cite{Bogol}, one is able
to add to the Lagrangean counterterms, which compensate the divergences.
But it is well known that this method is not universal --
there exist nonrenormalizable theories (which contain
infinite number of different types divergent Feynman
diagrams). An example is the theory of the weak
interactions, proposed by Fermi. Its
nonrenormalizability has led after long investigations to
the creation of the Weinberg-Sallam theory.

One possible way to overcome the problem with divergences in
QFT is to reject the restriction about the order of
derivatives in the Lagrangean (or in the equation of motion)
i.e., to consider Higher Derivative QFT (HDQFT). The main
advantage of HD theories is that propagators have better
behaviour at large momenta. That means that a theory containing a
certain number of divergences in the standard case, in the HD case
become (partially or completely) free from them.

Higher derivatives (HD) are not something new in field theory.
It is known since a long time that HD could be
used for regularization; different reguralarizations
\cite{Feynman} are equivalent to the usage of HD \cite{Iv-So}.
Regularization with higher derivatives is convenient to use in
the analysis of Yang-Mills fields \cite{Fadeev-Slav}.
HD have been used in classical electrodynamics \cite{Bopp-Pod}
for elimination of the problem with the infinite field--energy
of the electron. 

In the present paper we deal with a HDQFT for a spinor field
with scalar selfinteraction. The equation of motion for this
field is of third order in the derivatives. The coupling
constant is dimensionless (with respect to the term with
highest derivative), which is peculiar to the Yang-Mills
models. An important role in the model plays an additional
mass-parameter with large (but finite) value. This parameter
guarantees that the action is dimensionless and spinor fields have the 
standard dimension i.e. $3/2$. As we shall see below, the propagator 
in our model coincides with a spinor propagator regularized by 
Pauli-Villars, and the additional parameter plays the role of the 
regularization mass. That is why we suppose that this parameter is 
much larger than the mass of the physical field presented in the theory. 
After quantization, this parameter represents the mass of the spinor 
fields with nonpositive norm - the so called HD "ghosts" \cite{Hoking}.

In the nonrelativistic limit the equation of 
motion coincides with the non-linear Schr\"odinger equation. 
The basic hypothesis in our work is that (at least for low energies) 
the equation of motion has solutions of the Schr\"odinger type, 
i.e., two-particle bound state. The main goal of the present article is 
to verify this hypothesis. In order to do that, we use an analogue of 
the Bethe-Salpeter equation \cite{B.S.}.

\section{The model}\vskip 5mm

In the present paper we discuss a model of QFT with third order
(in the derivatives) Lagrangean
$L \sim \bar\varphi(\Box-M^2) (\partial_\mu \gamma^\mu -m)\varphi$.
However, for the derivation of the commutation relations and
Green functions it is more convenient to work with the following
(equivalent to the initial) second order Lagrangean
(with an additional spinor field $\psi$):
\bba
L&=&{1\over\sqrt{M^2-m^2}}(\pd^\mu \bar\psi
\pd_\mu\varphi + \pd^\mu \bar\varphi \pd_\mu\psi) +
\bar\varphi ({i\over 2}\ga^\mu\pd^{\!\!\!\!\! ^\leftrightarrow}_\mu
-
m)\varphi  \nonumber\\&& +
{\bar\psi}({i\over 2}\ga^\mu\pd^{\!\!\!\!\! ^\leftrightarrow}_\mu
+m)\psi - {m^2\over\sqrt{M^2-m^2}}({\bar\psi}\varphi +
{\bar\varphi}\psi) + {g\over{M^2-
m^2}}{({\bar\varphi}\varphi)}^2
\label{eq1}
\eea
where $A\pd^{\!\!\!\!\! ^\leftrightarrow}_\mu B=A\pd_\mu B-\pd_\mu AB$.
This Lagrangean contains an additional spinor field $\psi (x)$, and $M$
is the  arbitrary but finite mass parameter mentioned above.
The coupling constant $g$ is dimensionless and we
 shall assume it has arbitrary value and sign.

The equations of motion for both fields $\varphi$ and $\psi$
are:
\bba
{{(\Box+m^2)}\over\sqrt{M^2-m^2}}\psi&=&(i\ga^\mu\pd_\mu-
m)\varphi
+ {2g\over{M^2-m^2}}({\bar\varphi}\varphi)\varphi\nonumber\\
{{(\Box+m^2)}\over\sqrt{M^2-m^2}}\varphi&=&
(i\ga^\mu\pd_\mu+m)\psi
\label{eq3}
\eea
After exclusion of the auxiliary field $\psi$ from
(\ref{eq3}) we obtain a covariant equation of motion for the
physical field only:
\bb
(\Box+M^2)(i\ga^\mu\pd_\mu-m)\varphi +
2g({\bar\varphi}\varphi)\varphi=0
\label{eq4}
\ee
Let us consider its nonrelativistic limit. For this, it is
convenient to rewrite (\ref{eq4}) in the form:
$$\left( {\Box+m^2\over M^2-m^2}+1\right)
(i\ga^\mu\pd_\mu-m)\varphi(x)+
{2g\over M^2-m^2}(\bar\varphi\varphi) \varphi(x)=0$$
In momentum representation the first term has the form:
$$\left( {-p^2+m^2\over M^2-m^2} +1\right)
(\hat p -m)\varphi(p)$$
Because $M^2>>m^2$, we see, that for low momenta the term
${-p^2+m^2\over M^2-m^2}$ is extremely small. Consequently,
for low energies the behaviour of (\ref{eq4}) is determined
by the equation:
\bb
(i\ga^\mu\pd_\mu-m)\varphi+{2g\over M^2-
m^2}(\bar\varphi\varphi)\varphi=0
\label{eqLE}
\ee
This is a non-linear Dirac equation. Following the standard
procedure, it is easy to find its nonrelativistic limit:
\bb
\left( i{\pd\over \pd t} + {\Delta^2\over 2m}
\right)\varphi ' +
{2g\over M^2-m^2}(\bar\varphi '\varphi ')\varphi '=0
\label{eq4a}
\ee
Here $\varphi '$ is the large component of the spinor field
$\varphi$ with mass $m$. The equation (\ref{eq4a}) is
equivalent to the two-particle Schr\"odinger equation with
contact potential. The latter has the form:
\bb
i{\pd\over\pd t}\Psi(x_1,x_2;t)={\cal H}(x_1-
x_2)\Psi(x_1,x_2;t)
\label{eq5}
\ee
where:
$${\cal H}=-{1\over2m}\sum_{i=1}^2\nabla_i^2 +
\lambda\delta(x_1-x_2) = {\cal H}_0 + {\cal H}_{int}$$
Equation (\ref{eq5}) has a unique solution when the coupling
constant $\lambda$ has negative sign \cite{Berez.Fad}.
This suggests, that at least in the domain of low energies,
the equation (\ref{eq4}) could describe bound states of two
relativistic particles.

\section{Quantization of the model}\vskip 5mm

The quantization of the model is made in the interaction
representation. This procedure consists in the replacement of
the field functions with operators. We assume all operators
are normal ordered, and for simplicity we omit the sign for
normal ordering. The free equation of motion for the field $\varphi$
is:
\bb
{(\Box+M^2)\over{M^2-m^2}}(i\ga^\mu\pd_\mu-m)\varphi=0
\label{eq12}
\ee
The anticomutator
$\Gamma_{\alpha\beta}(x)=\{\varphi_\alpha(x),
{\bar\varphi}_\beta(0)\}$,
obeys the following initial conditions (see Appendix A):
\bb
\Gamma_{\alpha\beta}(x)\vert_{x_0=0}=0 \; ;\;
\pd_0\Gamma_{\alpha\beta}(x)\vert_{x_0=0}=0 \; ; \;
\pd_0^2\Gamma_{\alpha\beta}(x)\vert_{x_0=0}=(M^2-m^2)
\ga^0_{\alpha\beta}\delta^3(x)
\label{eq15}
\ee
It is obvious that $\Gamma(x)$ satisfy (\ref{eq12}). The
solution of (\ref{eq12}) with initial conditions
(\ref{eq15}) is:
\bb
\Gamma_{\alpha\beta}(x)=-i(i\ga^\mu\pd_\mu+m)_{\alpha\beta}
(D_m(x)-D_M(x))
\label{eq16}
\ee
where $D_{m}(x)$ is the Pauli-Jordan function for mass $m$.

The propagator of the field $\varphi$, here denoted by
${\cal S}^c_{\alpha\beta}(x)=
\left<  T(\varphi_\alpha(x){\bar\varphi}_\beta(0))\right> _0$,
is the solution of the equation:
\bb
{(\Box+M^2)\over{M^2-m^2}}(i\ga^\mu\pd_\mu-m)
i{\cal S}^c(x)=-\delta(x)
\label{eq17}
\ee
In momentum representation the solution of (\ref{eq17}) is:
$(k^\mu\ga_\mu+m)\left({1\over{m^2-k^2}}-
{1\over{M^2-k^2}}\right)$
and the propagator is:
\bb
{\cal S}^c_{\alpha\beta}(x)= -
i(i\ga^\mu\pd_\mu+m)_{\alpha\beta}
(D^c_m(x)-D^c_M(x))
\label{eq18}
\ee
where:
$$D^c_m(x)={1\over(2\pi)^4}\int{e^{-ikx}\over
{m^2-k^2-i\varepsilon}}d^4k$$
is the causal Green function of the scalar field. The
advantage of the usage of HD is obvious from the form (\ref
{eq18}) of the ${\cal S}^c$: the propagators in our model
coincide with the ones in the standard theory of a spinor field
regularized by Pauli-Villars. Now, it is easy to convince ourselves
that in the present theory there are no divergences connected to the
Feynman diagrams.

The solution of (\ref{eq12}) could be written as:
\bba
\varphi_\alpha(x)&=&\varphi_\alpha^{m}(x)\nonumber\\&&
+ {1\over (2\pi)^{3/2}}
\int d^3k \{ e^{ikx}(N_+u^{r,+}_\alpha(k)a^+_r(k) +
N_-v^{r,+}_\alpha(k)c^+_r(k)) \nonumber \\ && + e^{-ikx}(N_+
u^{r,-}_\alpha(k)a^-_r(k) + N_-v^{r,-}_\alpha(k)c^-_r(k)) \}
\label{eq22}
\eea
where $\varphi^{m}(x)$ is the usual spinor field satisfying
equation:
$(i\ga^\mu \pd_\mu -m)\varphi^{m}(x)=0$.
The mode-vectors $u(k)$ and $v(k)$ obey the equations:
\bba
(\mp k^\mu\ga_\mu-M)u^{r,\pm}(k)&=&0 \nonumber \\
(\mp k^\mu\ga_\mu+M)v^{r,\pm}(k)&=&0 \nonumber
\eea
and the usual normalising conditions
$${\bar u}^{r,\pm}(k)u^{s,\mp}(k)=\pm{M\over\Omega}
\delta^{r,s}={\bar v}^{r,\mp}(k)v^{s,\pm}(k)$$
\bba
\sum_su^{s,\pm}_\alpha(k){\bar u}^{s,\mp}_\beta(k)&=&
{(k^\mu\ga_\mu \mp M)_{\alpha\beta}\over2\Omega}
\nonumber \\
\sum_rv^{r,\pm}_\alpha(k){\bar v}^{r,\mp}_\beta(k)&=&
{(k^\mu\ga_\mu \pm M)_{\alpha\beta}\over2\Omega}
\label{eq21}
\eea
where
$\Omega=\sqrt{\vec{k}^2+M^2}$
and
$N_{\pm}=\sqrt{M \pm m\over2M}$.
The creation and annihilation operators $a^{\pm}$ and
$c^{\pm}$ in (\ref{eq22}), obey the following anticommutation relations:
\bba
\{a^-_r(k),a^{+^{\!\!\!\!\!\!\!\!\! *}}_s(p)\}&=&
\{a^{-^{\!\!\!\!\!\!\!\!\! *}}_r(k),a^+_s(p)\}=
-\delta_{rs}\delta(k-p) \nonumber \\
\{c^-_r(k),c^{+^{\!\!\!\!\!\!\!\!\! *}}_s(p)\}&=&
\{c^{-^{\!\!\!\!\!\!\!\!\! *}}_r(k),c^+_s(p)\}=
-\delta_{rs}\delta(k-p)
\label{eq25}
\eea
According to eq.(\ref{eq25}) these operators represent "ghosts"
(with mass $M$) in the model. The physical subspace is given
by vectors $\left.\mid ph\right> $, which satisfy the following
additional conditions:
\bba
\left. a^-\mid ph\right> &=&a^{-^{\!\!\!\!\!\!\!\!\! *}}\;
\left. \mid ph\right> =0
\nonumber \\
\left. c^-\mid ph\right> &=&c^{-^{\!\!\!\!\!\!\!\!\! *}}\;
\left. \mid ph\right> =0,
\nonumber
\eea

\section{The bound state equation}\vskip 5mm

In 1951 Bethe and Salpeter derived an relativistic integral
equation for the two-particle bound states \cite{B.S.}. With
the help of this equation, some relativistic corrections to
the fine and the superfine structure of the light atoms have
been calculated \cite{B.S.}, \cite{S.}. The kernel of this
integral equation consists of all two-particle irreducible
Feynman diagrams,
but in the case of a small coupling constant \cite{B.S.}
one can consider the kernel restricted to its first term
(the so called "ladder" approximation). Applying their approach,
we obtain an analogue equation for the interaction, considered in the paper.
Here, this approximation leads to extreme simplification -- due
to the presence of a  $\delta$-functions, the integral equation
is reduced to a set of $16$ linear algebraic equations. In this
approximation, the kernel corresponds to contact interaction
of two particles, and could be written as:
\bb
K(x_1,x_2,x_3,x_4)=-{ig\over{M^2-m^2}}\delta(x_1-x_3)
\delta(x_2-x_4)\delta(x_1-x_2)
\label{eq27}
\ee
Hence, the analogue of the Bethe-Salpeter equation is:
\bba
\chi_{\alpha_1\alpha_2}(p;x_1,x_2)&=&\int{G^0}_
{\alpha_1\alpha_2
\beta_1\beta_2}(x_1,x_2,x_3,x_4)K(x_3,x_4,x_5,x_6)
\times \nonumber \\
&&\chi_{\beta_1\beta_2}(p;x_5,x_6)dx_3dx_4dx_5dx_6
\label{eq32}
\eea
where $\chi(p;x_1,x_2)$ is the relativistic two-particle
wave function, defined as:
\bb
\chi_{\alpha\beta}(p;x_1,x_2)=\left< 0\mid
T(\varphi_{\alpha}(x_1)\varphi_{\beta}(x_2))
\mid p\right>
\label{eq30}
\ee
Here $p$ is the $4$-momenta of the system, and
$${G^0}_{\alpha_1\alpha_2\beta_1\beta_2}(x_1,x_2,x_3,x_4)=
{\cal S}^c_{\alpha_1\beta_1}(x_1-x_3){\cal
S}^c_{\alpha_2\beta_2}
(x_2-x_4)$$
is the product of two free Green functions (\ref{eq18})
corresponding to the two particles. Using the transformation
properties of the wave function:
$$\chi(p;x_1,x_2)=e^{ipa}\chi(p;x_1+a,x_2+a)$$
(where $a$ is an arbitrary $4$-vector) and introducing
absolute and relative co-ordinates $X$ and $x$ resp.:
\bba
X&=&{{x_1+x_2}\over2} \nonumber \\
x&=&x_1-x_2 \nonumber
\eea
we obtain: $\chi(p;x_1,x_2)=e^{-ipX}\chi(x)$. In the new
co-ordinates, instead of eq.(\ref{eq32}) we have:
\bb
\chi_{\alpha_1\alpha_2}(x)={ig\over(2\pi)^4{(M^2-m^2)}}
\int d^4k{\cal S}^c_{\alpha_1\beta_1}(k)
{\cal S}^c_{\alpha_2\beta_2}
(p-k)e^{-i(k-p/2)x}\chi_{\beta_1\beta_2}(x=0)
\label{eq34}
\ee

One can show, that in the nonrelativistic limit eq.
(\ref{eq34}) coincides with the Schr\"odinger equation with
$\delta$-function potential. Indeed, for low energies (see
(\ref{eqLE})) in the momentum representation one can rewrite
(\ref{eq34}) as:
$$\left({\hat p\over 2}+\hat q -m \right)_{\alpha_1\beta_1}
\left({\hat p\over 2}-\hat q -m\right)_{\alpha_2\beta_2}
\tilde\chi_{\beta_1\beta_2}(q)=-{ig\over M^2-m^2}
\chi_{\alpha_1\alpha_2}(x=0)$$
where $q$ is the relative momentum. Acting on both sides
of the above equation with the projectors
$\Lambda^{+}(\vec q)\Lambda^{-}(\vec q)$, where
$\Lambda^{\pm}(\vec q)={1\over2}+{(m\ga^0 \pm \vec q
\ga^0\vec{\ga})\over 2E(\vec q)}$
we obtain, after simple transformations:
$$\left({E\over2}-E(\vec q)+\varepsilon - i\delta\right)
\left({E\over2}-E(\vec q)-\varepsilon -i\delta\right)
\tilde \chi^{+,-}(\varepsilon,\vec q)=
-{i\lambda\over M^2-m^2}\chi^{+,-}(x=0)$$
Here $E$ is the energy in the rest frame $(\vec p=0)$,
$\varepsilon$ is the relative energy, $E(\vec q)=\sqrt{(\vec
q)^2 +m^2}$ and
$\tilde \chi^{+,-}(\varepsilon,{\vec q}) = \Lambda^{+}(\vec
q) \Lambda^{-}(\vec q) \tilde \chi(\varepsilon, {\vec q})$
(resp. for the function $\chi(0)$). The latter equation
could be integrated with respect to $\varepsilon$. The
result is:
$$\left(E-2E(\vec q)\right)
\int\tilde\chi^{+,-}(\varepsilon, \vec q) \; d\varepsilon=
{g\over M^2-m^2}\chi^{+,-}(x=0)$$
In the nonrelativistic limit the projectors could be removed
from the above (scalar) equation and we obtain exactly
the Schr\"odinger equation in the momentum representation:
$$(E-2m)\tilde\chi(\vec q)={q^2\over m}\tilde\chi(\vec q)
+{g\over M^2-m^2}\chi(x=0)$$

Let us return to the equation (\ref{eq34}). It is an
algebraic system connecting the wave function, taken
in different points ($x$ and $0$). Consequently, if we
set $x=0$ we obtain the following set of algebraic equations:
\bb
\chi_{\alpha_1\alpha_2}(x=0)={\cal F}_{\alpha_1\beta_1
\alpha_2 \beta_2}(p)\chi_{\beta_1\beta_2}(x=0)
\label{eq35}
\ee
where:
\bba
{\cal
F}_{\alpha_1\beta_1\alpha_2\beta_2}(p)&=&{ig\over(2\pi)^4
(M^2-m^2)}\int{\cal S}^c_{\alpha_1\beta_1}(k)
{\cal S}^c_{\alpha_2\beta_2}(p-k)d^4k \nonumber \\
{\cal S}^c(k)&=&-i(k^\mu\ga_\mu+m)
\left({1\over{m^2-k^2-i\varepsilon}}-
{1\over {M^2-k^2-i\varepsilon}}\right) \label{eq36}
\eea
The calculation of the integral (\ref{eq36}) is long, but it
is not connected to any principle difficulties. This
integral is similar to the vacuum polarisation diagram
in the Quantum Electrodynamics. Some details about the
calculation of such integrals one can see in \cite{Bogol}.

The final result can be written in the following form:
\bb
{\cal F}=a_{\mu\nu}\ga^\mu\otimes\ga^\nu+
b_\mu \left(1\!\!\!\:{\rm I}\otimes\ga^\mu+
\ga^\mu\otimes 1\!\!\!\:{\rm I}\right) +
d \; 1\!\!\!\:{\rm I}\otimes 1\!\!\!\:{\rm I}
\label{eq37}
\ee
where the functions $a_{\mu\nu} , b_\mu$ and $d$ in the rest
frame $p=(E,\vec 0)$ are given in Appendix B.
We are interested only in bound states between particles with
mass $m$. This means, that the values of the energy must be
in the interval $0<E^2<4m^2$. It is easy to check,  that in
this interval $a_{\mu\nu}, b_\mu, d $ are analytic
functions of the energy (the latter is not obvious only for
the point $E^2=0$).

If one writes the wave function $\chi$ as column with $16$
components, then the condition for compatibility of the
set of algebraic equations(\ref{eq35})
\bb
det({\cal F}_{16\times 16}- {1\!\!\!\:{\rm I}}_{16\times
16})=0
\label{eq39}
\ee
determines whether there exist bound states or not.

\section{Decomposition into Lorentz invariants}\vskip 5mm

For the subsequent analysis it is instructive to decompose
the two--particle wave function into Lorentz invariant
quantities. First, we start with the observation, (see
eq.(\ref{eq30})) that the quantity
$\tilde\chi(x)=\chi(x)\left(C^T\right)^{-1}$, where
$C$ is the charge conjugation matrix,
transforms as a tensor under bispinorial representation
$\tau({1\over2},0)\oplus\tau(0,{1 \over 2})$
of the Lorentz group, i.e.:
\bb
\tilde{\chi}'(x')=S^{-1}\tilde{\chi}(x) S
\;\;\;\; , \;\;\;\; x'_\mu=\Lambda_\mu^\nu x_\nu
\label{eq40}
\ee
This representation is reducible, and it decomposes into a
direct sum of irreducible representations: a scalar ($s$), a
pseudoscalar ($p$), a $4$-vector ($V_\alpha$), an axial
vector ($A_\alpha$) and an asymmetric $2$-rank tensor
($\Sigma_{\alpha\beta}$). Using the connection
$S^{-1}\ga^\mu S=\Lambda^\mu_\nu\ga^\nu, \;\;$
$\tilde{\chi}(x)$ could be written as:
\bb
\tilde{\chi}=s1\!\!\!\:{\rm I}+p\ga^5+V_\alpha\ga^\alpha+
A_\alpha\ga^5\ga^\alpha+\Sigma_{\alpha\beta}\sigma^{\alpha
\beta}
\label{eq41}
\ee
where
$\sigma^{\alpha\beta}={i\over 4}(\ga^\alpha\ga^\beta-
\ga^\beta\ga^\alpha)$.
After simple transformations, from (\ref{eq37}) and
(\ref{eq35}), one can obtain the equation for $\tilde\chi$:
\bb
a_{\mu\nu}\ga^\mu\tilde{\chi}\ga^\nu+b_\mu\left(\tilde{\chi}
\ga^\mu-\ga^\mu\tilde{\chi}\right) +(1-d)\tilde{\chi}=0
\label{eq44}
\ee
where we use that
$C^{-1}\ga_\mu C=-\ga_\mu^T$.
Finally, after standard $\ga$-matrix algebra, from
(\ref{eq44}) we obtain the following equations for the
Lorentz invariants:
\bba
(a_\mu^\mu-d+1)s&=&0\nonumber\\
(a_\mu^\mu+d-1)p&=&2b_\mu A^\mu\nonumber\\
2a_\mu^\sigma V^\mu -(a_\mu^\mu+d-1)V^\sigma &=&2b_\mu i
\Sigma^{\mu \sigma}
\label{eq46}\\
2a_\mu^\sigma A^\mu-(a_\mu^\mu-d+1)A^\sigma &=&2b^\sigma p
\nonumber\\
(a_\mu^\mu-d+1)i\Sigma^{\sigma\rho}-
2(a^{\sigma\alpha}i\Sigma_\alpha^\rho
-a^{\rho\alpha}i\Sigma_\alpha^\sigma)&=&2(b^\sigma V^\rho-
b^\rho V^\sigma)\nonumber
\eea
Hence, we see that in terms of fields $s, p, V, A, \Sigma$
the matrix ${\cal F}_{16\times 16}$ could be written as:
$${\cal F}_{16\times 16}={\cal F}_{1\times 1}(s)\oplus
{\cal F}_{5\times 5}(p,A)\oplus{\cal F}_{10\times 10}
(V,\Sigma)$$
Therefore, the condition for existing of bound states
(\ref{eq39}) decomposes to tree independent conditions:
\bb
{\cal F}(s) - 1 =0 \;\;,\;\; \det({\cal F}(p,A)-1\!\!\!\:
{\rm I})=0 \;\;,\;\; det({\cal F}(V,\Sigma)-1\!\!\!\:
{\rm I})=0
\label{eq47}
\ee

\section{Study of the scalar bound state}\vskip 5mm

Let us concentrate our attention on the conditions
(\ref{eq47}). It is clear that the condition for the scalar
bound state is the simplest one, because the scalar field
does not mix with the other fields. In this section we solve
the condition for the scalar bound state, which (following
(\ref{eq46})) is:
\bb
a_\mu^\mu-d+1=0
\label{eq48}
\ee
This equation is too complicated to be solved in the general
case. But since we have a small parameter - $E^2/M^2$ - we
can simplify eq.(\ref{eq48}). For this purpose, we use the
approximation of the functions $a_{\mu\nu}$ and $d$ given
into Appendix B. Then eq.(\ref{eq48}) is:
\bba
1&+&{g\over 16\pi^2}{M^2\over M^2-m^2} +{g\over
16\pi^2}{m^2\over M^2-m^2}\left\{\left({E^2\over 2m^2}-
3\right) ln\left({M^2\over m^2}\right)
\right.\nonumber\\
&&\left. -1+ \left(4-{E^2\over m^2}\right) {R_m\over E}
arctg\left({E\over R_m}\right)\right\}
+O\left({E^4\over M^4}\right)=0
\label{eq49}
\eea
An interesting feature of this equation is that it contains
two types of terms: the first two are of the order of $1$,
while the rest are $\sim m^2/M^2$ and therefore are much
smaller than the first two. This means that (\ref{eq49})
has solution if the large and small terms are
equal to zero independently i.e. (\ref{eq49}) decomposes
in two different conditions for the large and small parts.
These two equations are:
\bba
&&{g'\over 16\pi^2}+1=0\label{eq50}\\
&&\Delta g+\left({E^2\over 2m^2}-3\right)
ln\left({M^2\over m^2}\right)-1 \nonumber\\
&&+\left(4-{E^2\over m^2}\right) {R_m\over E}
arctg\left({E\over R_m}\right)=0
\label{eq50.1}
\eea
\underline{Remark}: The separation of the eq.(\ref{eq49})
in two equations is possible, if the large part is
determined up to term of order $m^2/M^2$. This arbitrary
term contributes in the equation for the small parts. In
order to do that we represent t
he coupling constant as:
$g=g'\left(1+\Delta g {m^2\over M^2}\right)$.
From (\ref{eq50}) we have:
$g'=-16\pi^2$.
It is easy to verify that (\ref{eq50.1}) has a unique solution
which depends on $M^2$ and $\Delta g$. Indeed, the LHS of
(\ref{eq50.1}) is a monotonously increasing function of $E$,
and at the ends of the interval
$0\leq E^2 \leq 4m^2$
it has the values:

$$LHS (\ref{eq50.1}) =\left\{
\begin{array}{ll}
\Delta g +3-3ln\left({M^2\over m^2}\right)    &E^2=0\\
\Delta g -1-ln\left({M^2\over m^2}\right)    &E^2=4m^2
\end{array}
\right.$$
Consequently, the model considered here admits the existence
of the scalar bound state with mass (in the rest frame)
$E^2(\Delta g, M^2)$, which is the solution of eq.(\ref{eq50.1}).
This bound state exists for each $M^2: M^2>>m^2$, when the
coupling constant is negative and has the value:
$$g=-16\pi^2\left(1+\Delta g {m^2\over M^2}\right)$$
where: $\;\;\;    ln\left({M^2\over m^2}\right)+1<\Delta g<
3ln\left({M^2\over m^2}\right) -3$.
This result is also in correspondence with the solution of
the non-linear Schr\"odinger equation, discussed above.

In conclusion, we want to mark the possibility for solutions
of (\ref{eq47}), connected to the other fields.
It is easy to verify from eq.(\ref{eq46}), that in the rest
frame, the matrices ${\cal F}(p,A)$ and ${\cal
F}(V,\Sigma)$, has only symmetrical placed nondiagonal terms
(besides the main diagonal) which are $\sim m^2/M^2$.
Consequently, their contribution to the determinant is $\sim
m^4/M^4$ or smaller. Because of this, we can consider
these nondiagonal elements in these matrices as zeros.
Finally we obtain equations of the type (\ref{eq49}),
for the four fields in consideration. To solve these
equations we proceed in analogy with (
\ref{eq50}) and (\ref{eq50.1}). In order to satisfy the
(corresponding) condition for the large parts, we have to
choose the constant $g'$ to be:
$$ g'_p =16\pi^2 \;\; , \;\;
   g'_V =32\pi^2 \;\; , \;\;
   g'_A =-32\pi^2 $$
The root, corresponding to $\Sigma$ field is not achievable,
because the large component in this case is zero. It
is important to note, that equations for small parts,
corresponding to temporal and spatial components of the
vector and axial-vector fields are different.

\vskip 6mm \centerline{\bf Appendix A}\vskip 8mm

First, we introduce two auxiliary functions
$\Delta_{\alpha\beta}(x)=\{\psi_\alpha(x),{\bar\psi}_\beta(0
)\}$
and
$S_{\alpha\beta}(x)=\{\varphi_\alpha (x),\bar\psi_\beta
(0)\}$.

From the requirement that fields, separated with space-like
interval are causal independent, it follows, that:
\bb
\Gamma_{\alpha\beta}(x)\vert_{x^0=0}=
\Delta_{\alpha\beta}(x)\vert_{x^0=0}=
S_{\alpha\beta}(x)\vert_{x^0=0}=0
\label{p1.1}
\ee
The procedure of canonical quantization gives:
\bb
\{\pi_{\varphi_\alpha}(x),\varphi_\beta(0)\}\vert_{x^0=0}=
\{\pi_{\psi_\alpha}(x),\psi_\beta(0)\}\vert_{x^0=0}
=i\delta^3(x)\delta_{\alpha\beta}
\label{p1.2}
\ee
where $\pi$ is canonical conjugated momentum for
corresponding field. Substituting in (\ref{p1.2}) the
explicit form of conjugated momenta, we have:
\bb
\pd_0\Gamma_{\alpha\beta}(x)\vert_{x^0=0}=
\pd_0\Delta_{\alpha\beta}(x)\vert_{x^0=0}=0
\label{p1.4}
\ee
and
\bb
\pd_0 S_{\alpha\beta}(x)\vert_{x^0=0}=
-i\sqrt{M^2-m^2}\delta^3(x)\delta_{\alpha\beta}
\label{p1.5}
\ee
Let us consider eq.(\ref{eq3}) in the free case. One can
show, that:
$${1\over\sqrt{M^2-
m^2}}\pd^2_0\Gamma_{\alpha\beta}(x)\vert_{x^0=0}=i\ga^0_{\alpha
\rho}\pd_0 \{\psi_\rho (x),
\bar\varphi_\beta(0)\}\vert_{x^0=0}$$
Finally, from the latter equation and from eq.(\ref{p1.5}) we
obtain:
\bb
\pd_0^2\Gamma_{\alpha\beta}(x)\vert_{x^0=0}=
(M^2-m^2)\ga^0_{\alpha\beta}\delta^3(x)
\label{p1.7}
\ee

\vskip 6mm\centerline{\bf Appendix B}\vskip 8mm

In the rest frame $p=(E,\vec 0)$, functions $a_{\mu\nu},
b_\mu$ and $d$ in (\ref{eq37}) are:
\bba
a_{\mu\nu}&=&{g\over16\pi^2 E^2(M^2-m^2)}\left\{
{g_{\mu\nu}\over 2} \left[-{(M^2-m^2)^2\over 3}
\right .\right .\nonumber\\
&&+(3E^4-3E^2(M^2+m^2)+(M^2-m^2)^2){(M^2-
m^2)\over6E^2}ln\left({M^2\over m^2}\right)\nonumber\\
&&-{1\over 3E^2} R^3 \; arth\left( {R\over M^2+m^2-E^2}
\right)
\nonumber\\
&&\left.+{E\over 3}R^3_m arctg\left({E\over
R_m}\right)\;\;\; + \;\;\;{E\over 3}R^3_M
arctg\left({E\over R_M}\right) \right]
\nonumber\\
%
&&+g_{\mu 0}g_{\nu 0}\left[{2\over3}(M^2-m^2)^2    \right.
\nonumber\\
&&+(3E^2(M^2+m^2)-2(M^2-m^2)^2){(M^2-
m^2)\over6E^2}ln\left({M^2\over m^2}\right)\nonumber\\
&&-{1\over 3E^2}(E^4+E^2(M^2+m^2)-2(M^2-
m^2)^2)R\;\;arth\left({R\over M^2+m^2-E^2}\right)\nonumber\\
&& \left.\left.-{E\over 3}(2m^2+E^2)R_m
arctg\left({E\over R_m}\right)\;\;\; - \;\;\;{E\over
3}(2M^2+E^2)R_M  arctg\left({E\over R_M}\right) \right]
\right\} \nonumber
\eea
\bba
d&=&{2m^2g\over 16\pi^2E^2(M^2-m^2)}\left\{ {(M^2-m^2)
\over 2}ln\left({M^2\over m^2}\right)-R\;\;
arth\left({R\over M^2+m^2-E^2}
\right) \right.\nonumber\\
&&\left. -ER_m  arctg\left({E\over R_m}\right)-ER_M
arctg\left({E\over R_M}\right)
\right\}\nonumber\\
b_\mu&=&{E\over 2m}g_{\mu 0} d\nonumber
\eea
where:
$$R=\sqrt{(E^2-M^2-m^2)^2-4M^2 m^2}\;\;\; R_m=\sqrt{4m^2-
E^2}
\;\;\;R_M=\sqrt{4M^2-E^2}$$
If we recall that $E^2/M^2$ is an extremely small parameter,
then the above functions could be rewritten in the form:
\bba
a_{\mu\nu}&=&{g\over16\pi^2 (M^2-m^2)}\left\{
g_{\mu\nu}\left[
{M^2\over 4}+{m^2\over 2}\left[ -ln\left({M^2\over
m^2}\right)-{5\over 6}\right.\right.\right.\nonumber\\
&&\left.\left. +\left({1\over 6}ln\left({M^2\over
m^2}\right)-{1\over18}\right)
{E^2\over m^2}+{1\over 3}\left(4-{E^2\over m^2}\right)
{R_m\over E} arctg\left({E\over R_m}\right) \right] \right]
\nonumber\\
&&\left.+ g_{\mu 0}g_{\nu 0} m^2\left[{2\over 3}
+\left({1\over 6}ln\left({M^2\over m^2}\right)+{1\over
9}\right){E^2\over m^2}-{1\over 3}\left(2+{E^2\over
m^2}\right){R_m\over E}arctg\left({E\over R_m}\right)
\right]\right\} \nonumber \\
&& + O\left({E^4\over M^4}\right)\nonumber\\
d&=&{g\over 16\pi^2(M^2-m^2)} \;m^2\left\{
ln\left({M^2\over m^2}\right) -2{R_m\over E} arctg
\left({E\over R_m}\right)\right\}+O\left({E^4\over
M^4}\right)\nonumber
\eea


\begin{thebibliography}{99}
%
\bibitem{Bogol} N.N.Bogolubov and D.V.Shirkov,
{\it Introduction in Quantum Field Theory}
(Moscow, Nauka, 1973).
%
\bibitem{Feynman}  R.P. Feynman, {\it Phys. Rev.}
{\bf 74}, 939 (1948) ; R.P. Feynman, {\it Phys. Rev.}
{\bf 76}, 769 (1949)
%
\bibitem{Iv-So} D. Ivanenko and V. Grigoriev, {\it JETP}
{\bf 21}, 563 (1951) (in russian)
%
\bibitem{Fadeev-Slav} L.D. Fadeev and A.A. Slavnov,
{\it Gauge fields: an introduction to quantum theory}
(Addison-Wesley Publishing Company, 1991)
%
\bibitem{Bopp-Pod} F. Bopp, {\it Ann. d. Phys.} {\bf 38},
345
(1940) ; B. Podolsky, {\it Phys. Rev.} {\bf 62}, 68 (1941) 
%
\bibitem{Hoking} S. Hawking, {\it Who's Afraid of (Higher
Derivative)
Ghosts}? in {\it Quantum Field Theory and Quantum
Statistics} {\bf 2}, 129 (Cambridge, 1985)
%
\bibitem{B.S.} E.E. Salpeter and H.A. Bethe,
{\it Phys.rev.} {\bf84}, 1232 (1951) 
%
\bibitem{S.} E.E. Salpeter, {\it Phys.rev.} {\bf 87},
328 (1952)  
%
\bibitem{Berez.Fad} F.A. Berezin and L.D. Fadeev,
{\it Soviet
Math. Dokl.} {\bf 2}, 372 (1961) ; S. Albeverio,
F. Gesztesy, R. Hoegh-Krohn and H. Holden,
{\it Solvable models in Quantum Mechanics}
(Springer-Verlag, 1988)
%


\end{thebibliography}
\end{document}